\documentclass[11pt]{article}

\usepackage{amssymb,amsmath,latexsym, hyperref, cite, natbib, graphicx, refstyle, pgfplotstable, booktabs, rotating, color, tabularx, authblk, array, url, soul, longtable,lscape,afterpage,cclicenses}

\usepackage{multirow}
\newcolumntype{C}[1]{>{\centering\arraybackslash}p{#1}}

\usepackage{geometry}
\usepackage{setspace}
\usepackage[multiple, hang,splitrule]{footmisc}


\usepackage[font=small, margin=0cm]{caption}

\begin{document}
\title{\textbf{What Is an Emerging Technology?}}
\author[1,2]{\textbf{Daniele Rotolo}\thanks{Corresponding author: d.rotolo@sussex.ac.uk, Phone: +44 1273 872980}}
\author[2]{\textbf{Diana Hicks}\thanks{diana.hicks@pubpolicy.gatech.edu}}
\author[1,3]{\textbf{Ben R. Martin}\thanks{b.martin@sussex.ac.uk}}

\affil[1]{\small SPRU --- Science Policy Research Unit, University of Sussex, Brighton, United Kingdom}
\affil[2]{\small School of Public Policy, Georgia Institute of Technology, Atlanta, United States}
\affil[3]{\small Centre for Science and Policy (CSAP) and Centre for Business Research, Judge Business School, University of Cambridge, Cambridge, United Kingdom}

\date{Version: \today \linebreak 
Accepted for publication in \textit{\textbf{Research Policy}}\thanks{\href{http://dx.doi.org/10.1016/j.respol.2015.06.006}{{\color{blue}DOI: 10.1016/j.respol.2015.06.006}}. \copyright 2015 Rotolo, Hicks, Martin.  Distributed under \href{http://creativecommons.org/licenses/by/4.0/}{{\color{blue}CC-BY-NC-ND}}.}}

%-------------------------------------------------------------------------------
%	Abstract
%-------------------------------------------------------------------------------
\maketitle
\begin{abstract}
\onehalfspacing
\noindent There is considerable and growing interest in the emergence of novel technologies, especially from the policy-making perspective. Yet as an area of study, emerging technologies lacks key foundational elements, namely a consensus on what classifies a technology as 'emergent' and strong research designs that operationalize central theoretical concepts. The present paper aims to fill this gap by developing a definition of 'emerging technologies' and linking this conceptual effort with the development of a framework for the operationalisation of technological emergence. The definition is developed by combining a basic understanding of the term and in particular the concept of 'emergence' with a review of key innovation studies dealing with definitional issues of technological emergence. The resulting definition identifies five attributes that feature in the emergence of novel technologies. These are: (i) radical novelty, (ii) relatively fast growth, (iii) coherence, (iv) prominent impact, and (v) uncertainty and ambiguity. The framework for operationalising emerging technologies is then elaborated on the basis of the proposed attributes. To do so, we identify and review major empirical approaches (mainly in, although not limited to, the scientometric domain) for the detection and study of emerging technologies (these include indicators and trend analysis, citation analysis, co-word analysis, overlay mapping, and combinations thereof) and elaborate on how these can be used to operationalise the different attributes of emergence.  \newline\newline
\par
\end{abstract}
\clearpage 

%-------------------------------------------------------------------------------
%	Introduction
%-------------------------------------------------------------------------------
\section{Introduction}
\begin{figure}[h]
\includegraphics[width=16cm]{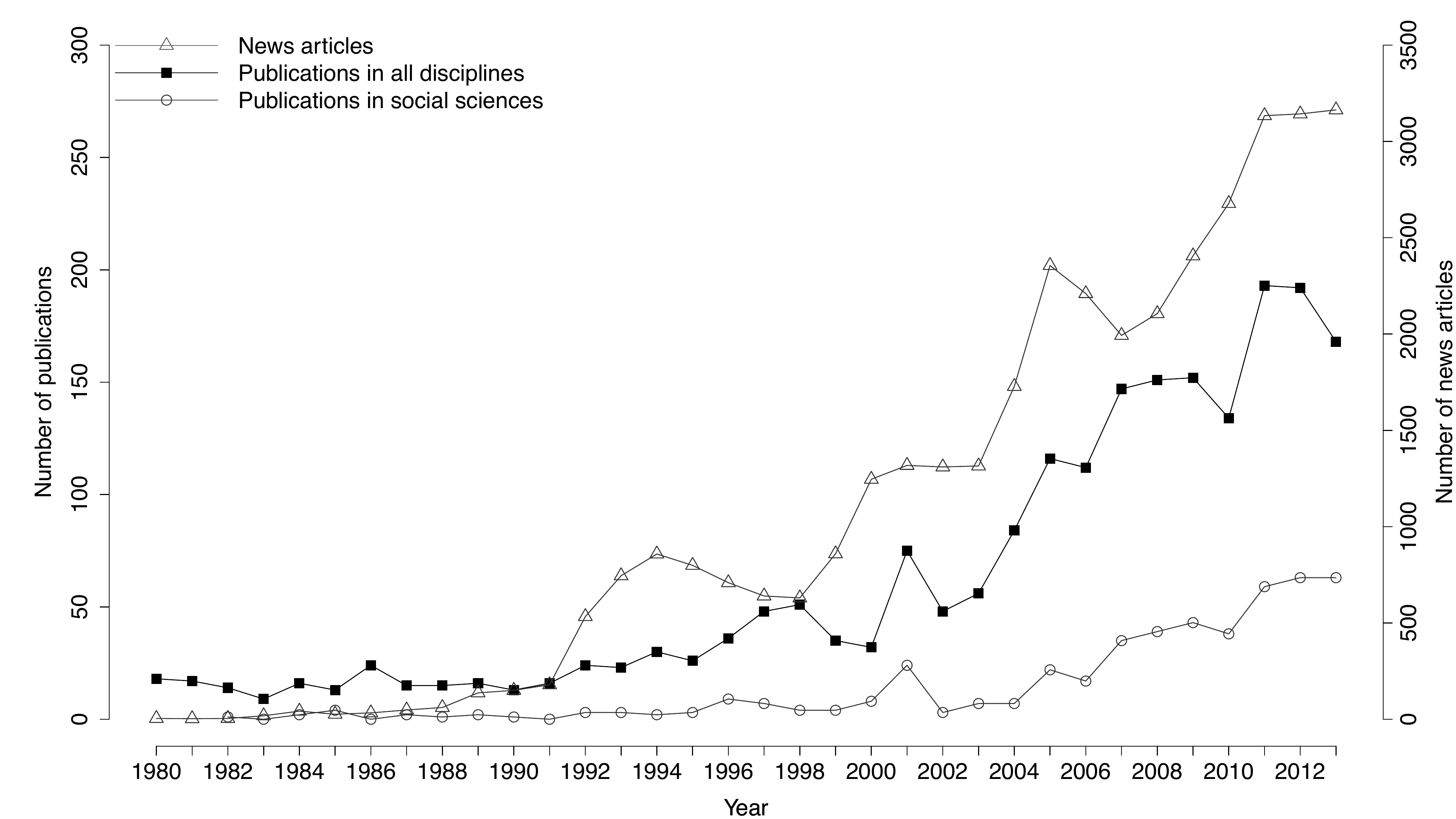}
\centering
\caption{Publications (left axis) and news articles (right axis) including the variations of the term "emerging technologies". Publications were retrieved by querying SCOPUS data: \textit{"TITLE("emerg* technol*") OR TITLE("emergence of* technolog*") OR TITLE("techn* emergence") OR TITLE("emerg* scien* technol*")"}. Publications in social sciences were defined as those assigned to the SCOPUS categories "Business, Management and Accounting", "Decision Sciences", "Economics, Econometrics and Finance", "Multidisciplinary", "Psychology", and "Social Sciences". News articles were identified by searching for "\textit{emerg* near2 technolog*}" in article headlines and lead paragraphs as reported in FACTIVA. From 1980 to 2013, the average yearly growth rates of the number of publications concerning emerging technologies in all disciplines and in social sciences have been of 12.5\% and 23.8\%, respectively. The total number of publications in SCOPUS has yearly grown on average by 4.9\%. \newline\textit{Source: search performed by authors on SCOPUS and FACTIVA.}}
\label{fig:scopus}
\end{figure}

Emerging technologies have been the subject of much debate in academic research and a central topic in policy discussions and initiatives. Evidence of the increasing attention being paid to the phenomenon of emerging technologies can be found in the growing number of publications dealing with the topic and news articles mentioning emerging technologies (in their headlines or lead paragraphs), as depicted in \Figref{scopus}. Increasing policy interest in emerging technologies, however, must be set against a literature where no consensus has emerged as to what qualifies a technology to be emergent. Definitions proposed by a number of studies overlap, but also point to different characteristics. For example, certain definitions emphasise the potential impact emerging technologies are capable of exerting on the economy and society \citep[e.g.][]{Porter2002}, especially when they are of a more 'generic' nature \citep{Martin1995}, while others give great importance to the uncertainty associated with the emergence process \citep[e.g.][]{Boon2008} or to the characteristics of novelty and growth \citep[e.g.][]{Small2014}. The understanding of emerging technologies also depends on the analyst's perspective. An analyst may consider a technology emergent because of its novelty and expected socio-economic impact, while others may see the same technology as a natural extension of an existing technology. Also, emerging technologies are often grouped together under 'general labels' (e.g.\ nanotechnology, synthetic biology), when they might be better treated separately given their different socio-technical features (e.g.\ technical difficulties, involved actors, applications, uncertainties).

The lack of consensus over definitions is matched by an 'eclectic' and \textit{ad hoc} approach to measurement.  A wide variety of methodological approaches have been developed, especially by the scientometric community, for the detection and analysis of emergence in science and technology domains \citep[e.g.][]{Porter1995, Boyack2014, Glanzel2012}. These methods, favoured, because they take advantage of growing computational power and large new datasets and allow one to work with more sophisticated indicators and models, lack strong connections to well thought out concepts that one is attempting to measure, a basic tenet of good research design. Often no definition of the central concept of an emerging technology is provided. It is no surprise therefore that approaches to the detection and analysis of emergence tend to differ greatly even with the use of the same or similar methods. The operationalisation of emergence is also in a state of flux. It changes as new categorisations (e.g.\ new terms in institutionalised vocabularies, new technological classes) are created within databases. This, in turn, makes less clear the exact nature of the phenomena that these scientometric methods enable us to examine. 

These problems in the effort to understand emerging technologies limit the utility of the research and so may hamper resource allocation and the development of regulations, which, in turn, have a major role in supporting and shaping the directionality of technological emergence.

%=====
%=====

The present paper addresses both the conceptual and methodological gaps. We aim to elaborate a framework that links what is conceptualised as 'emerging technologies' with its measurement, thus providing guidance to future research (e.g.\ development of novel methods for the detection of emergence and analysis of its characteristics) and to policy-making (e.g.\ resource allocation, regulation). To do so, we first attempt to clarify the conceptualisation of emerging technologies by integrating different conceptual contributions on the topic into a more precise and coherent definition of 'emerging technology'. We begin with the definition of 'emergence' or 'emergent', which is the process of coming into being, or of becoming important and prominent. This is then enriched and contextualised with a review of major contributions to innovation studies that have focused on technological emergence, highlighting both their common and contradictory features. Conceptual attempts to grapple with emergence in complex systems theory are also discussed where relevant to the idea of emergent technology.  

The result is the delineation of five key attributes that qualify a technology as emerging. These are: (i) radical novelty, (ii) relatively fast growth, (iii) coherence, (iv) prominent impact, and (v) uncertainty and ambiguity. Specifically, we conceive of an emerging technology as \textit{a radically novel and relatively fast growing technology characterised by a certain degree of coherence persisting over time and with the potential to exert a considerable impact on the socio-economic domain(s) which is observed in terms of the composition of actors, institutions and patterns of interactions among those, along with the associated knowledge production processes. Its most prominent impact, however, lies in the future and so in the emergence phase is still somewhat uncertain and ambiguous}. 

Second, the framework for operationalising emerging technologies is developed on the basis of the attributes we identified. The scientometric literature forms the core of the methods discussed because, as mentioned, this field has been remarkably active in developing methodologies for the detection and analysis of emergence in science and technology. The reviewed methods are grouped into five main categories: (i) indicators and trend analysis, (ii) citation analysis (including direct citation and co-citation analysis, and bibliographic coupling), (iii) co-word analysis, (iv) overlay mapping, and (v) hybrid approaches that combine two or more of the above. Because scientometric techniques cannot address all the attributes comprehensively, we also discuss approaches developed in other fields. 

The paper is organised as follows. The next section introduces the concept of emergence and its various components. In Section 3, these elements are integrated with key innovation studies proposing definitions of technological emergence, and a definition of emerging technologies is then elaborated. Section 4 reviews methods to both detect and analyse emergence, and then examines the use of those approaches to operationalise the proposed definition and the various attributes of emerging technologies. Section 5 discusses the limits of current methodologies for the detection and analysis of emerging technologies and identifies directions for future research. Section 6 summarises the main conclusions of the study.

%-------------------------------------------------------------------------------
%	The concept of emergence
%-------------------------------------------------------------------------------
\section{The concept of emergence}

The word 'emerge' or 'emergent' means "the process of coming into being, or of becoming important and prominent" (New Oxford American Dictionary) or "to rise up or come forth [...] to become evident [...] to come into existence" (the American Heritage Desk Dictionary and Thesaurus). \Tabref{dictionary} presents dictionary definitions of emergent. The primary attribute of emergence is 'becoming' --- that is, coming into existence.  Emergent is not a static property; it is a label for a process.  The endpoint of the process is variously described as visible, evident, important or prominent.  Thus, among the dictionaries there is some disagreement as to whether acknowledged existence is enough for emergence, or beyond that, a certain level of prominence is needed in order to merit application of the term emergence.

%=====
\setlength{\tabcolsep}{10pt}
\renewcommand{\arraystretch}{1.5}
\begin{table}\footnotesize
	\caption{\label{tab:dictionary}Dictionary definitions of the concept of emergence.}
	\centering
{\begin{tabular}{p{10cm}p{5cm}}
\hline\hline
\textbf{Dictionary definition of 'emerge'/'emergent'}& \textbf{Attributes}\\
\hline
"the process of coming into being, or of becoming important and prominent" (New Oxford American Dictionary) & come into being; important; prominent\\ 

"to become manifest: become known [...]" (Merriam-Webster's Collegiate Dictionary) & become manifest; become known\\
     
"to rise up or come forth [...] to become evident [...] to come into existence" (The American Heritage Desk Dictionary and Thesaurus) & evident; come into existence\\
          
"move out of something and become visible [...] come into existence or greater prominence [...] become known [...] in the process of coming into being or prominence" (Concise Oxford English Dictionary) & visible; prominent; become known; come into being\\

"starting to exist or to become known [...] to appear by coming out of something or out from behind something (Cambridge Dictionaries Online) & become known; to appear\\

\hline\hline
\multicolumn{2}{l}{\footnotesize \textit{Source: search performed by authors on major English dictionaries.}}
\end{tabular}
}
\end{table}
%=====

There is a second definition of emergent given the by the New Oxford American Dictionary as: a property arising as an effect of complex causes and not analysable simply as the sum of their effects. An additional definition is: arising and existing only as a phenomenon of independent parts working together, and not predictable on the basis of their properties. This concept of emergence is used in the study of complex systems. It can be traced back to the 19th Century in the proto-emergentism movement when Lewes (\citeyear{Lewes1875}) referred to 'emergent effects' in chemical reactions as those effects that cannot be reduced to the components of the system, i.e.\ the effects for which it is not possible to trace all the steps of the processes that produced them. Its application in the study of the dynamics of complex systems in physics, mathematics, and computer science gave rise to other fundamental theories and schools of thought such as complex adaptive system theory, non-linear dynamical system theory, the synergetics school, and far-from-equilibrium thermodynamics \citep[see][]{Goldstein1999}. 

A number of studies focusing on the definitional issue of emergence were produced by scholars in complex system theory --- see \Tabref{emergence} in the Appendix for an overview of the definitions of emergence proposed by major studies in complex system theory. Goldstein (\citeyear{Goldstein1999}), for example, defined emergence as "the arising of novel and coherent structures, patterns, and properties during the process of self-organization in complex systems" (\citeyear[p. 49]{Goldstein1999}). An ontological and epistemological definition of emergence is instead developed by \cite{DeHaan2006}. Ontological emergence is "about the properties of wholes compared to those of their parts, about systems having properties that their objects in isolation do not have" (\citeyear[p. 294]{DeHaan2006}), while epistemological emergence it is about "the interactions between the objects that cause the coming into being of those properties, in short the mechanisms producing novelty" (\citeyear[p. 294]{DeHaan2006}). 

Though research on complex systems may have a certain cachet (and perhaps for this reason scholars of emerging technologies sometimes attempt to work with the meaning of emergent as conceived by the complex system approach), we maintain that questions about emerging technologies are not fundamentally about understanding the origins and the causal nature of full system interaction; rather they are about uncertainty, novelty, identification at an early stage, and visibility and prominence. It is true that some technologies in themselves may be complex systems in the sense of exhibiting adaptation, self-organisation, and emergence, an example being parts of materials science \citep{Ivanova1998}. However, other technologies exhibit 'complicatedness' rather than 'complexity' as defined in complex system theory --- for example, engineering systems. These systems are designed for specific purposes, but they do not adapt and self-organise to changes in the environment \citep{Ottino2004}. It is also true that emerging technologies may arise from complex innovation systems \citep{Katz2006}, but we would contend that in the phrase 'emerging technology', 'emerging' is generally understood in the standard sense, not the complex system usage.

%-------------------------------------------------------------------------------
%	Defining emerging technologies
%-------------------------------------------------------------------------------
\section{Defining emerging technologies}
To further clarify what is meant by emerging technology, we reviewed literature in innovation studies dealing with definitional issues of emerging technologies. To identify relevant studies, we searched for \textit{"emerg* technolog*"}, \textit{"tech* emergence"}, \textit{"emergence of* technolog*"},  or  \textit{"emerg* scien* technol*"} in publication titles by querying SCOPUS (see the left-hand column of \Tabref{searches}).\footnote{The terminology of 'emerging technologies' has become central to a number of research traditions and especially to the scientometric, bibliometric and tech-mining domains \citep[cf.][]{Avila2011}, which, as discussed, have been remarkably active in developing methods for the operationalisation of emergence. In other words, 'emerging technologies' have become a category of its own. For this reason, we do not include epistemologically related terms, such as 'radical', 'disruptive', 'discontinuous', 'nascent' and 'breakthrough'.} We restricted the search to the title field to limit results to publications primarily focused on emerging technologies. The search identified a total of 2,201 publications from 1971 to mid 2014.\footnote{The search was performed on 13th May 2014.} Within this sample we selected those publications in social science domains, thus reducing the sample to 501 records (see \Figref{scopus}). 

We then read the abstracts and accessed the full-text of these studies where necessary both to identify additional documents from the list of cited references and to exclude studies that are not relevant to the scope of this paper.  We found that about 50\% of the studies in the sample refer to a specific industrial context (e.g.\ listing and discussing emerging technologies in a given industry) or to the educational sector (e.g.\ emergence of novel technologies to improve education and learning). These were deemed not relevant to our study. The remaining studies were further examined to identify those that develop or provide definitions of emerging technologies --- we searched for 'defining' sentences within the publication full-text by using the keywords listed above. This led to a core set of 12 studies from science and technology (S\&T) policy studies, evolutionary economics, management, and scientometrics that contributed to the conceptualisation of technological emergence. These are listed with their definitions of emerging technologies in \Tabref{techemergence}. We analysed the textual content of the definitions reported in \Tabref{techemergence} to extract all the component concepts. These were grouped into the attributes discussed below and used to construct our definition of emerging technologies. Extracted concepts excluded from our list of attributes will also be discussed.

%=====
\setlength{\tabcolsep}{10pt}
\renewcommand{\arraystretch}{1.5}
\begin{table}[h]\footnotesize
	\caption{\label{tab:searches}Search strategies used to identify the set of relevant publications for the conceptualisation and operationalisation of emerging technologies.}
	\centering
{\begin{tabular}{lp{4cm}p{6cm}}
\hline\hline
\textbf{ }&	\textbf{Conceptualisation} & 							\textbf{Operationalisation} \\
\hline
\textbf{Search terms }&		\textit{"emerg* technolog*"} & 								\textit{"emerg* technolog*"} \\
  &						\textit{"tech* emergence"} &								\textit{"tech* emergence"} \\
  &						\textit{"emergence of* technolog*"} &							\textit{"emergence of* technolog*"} \\
  &						 \textit{"emerg* scien* technol*"} &							\textit{"emerg* scien* technol*"} \\
  &												&							\textit{"emerg* topic*"} \\
  &												&							\textit{"emergence of* topic*"} \\
\textbf{Field(s) of search} 		&	Title												& Title, abstract, keywords	 \\
\textbf{Focus} &	Social sciences	& Scientometric journals: \textit{Journal of the Association for Information Science \& Technology} (formerly the \textit{Journal of the American Society for Information Science \& Technology}), \textit{Journal of Informetrics}, \textit{Research Evaluation}, \textit{Research Policy}, \textit{Scientometrics}, \textit{Technological Forecasting \& Social Change}, \textit{Technology Analysis \& Strategic Management}\\
\textbf{Number of studies} 	&	501					&							155\\
\hline

\hline\hline
\multicolumn{3}{l}{\footnotesize \textit{Source: authors' elaboration as based on SCOPUS data.}}
\end{tabular}
}
\end{table}
%=====

%=====
\setlength{\tabcolsep}{10pt}
\renewcommand{\arraystretch}{1.4}
\begin{table}\footnotesize
	\caption{\label{tab:techemergence}Definitions of emerging technologies (studies are chronologically ordered).}
	\centering
{\begin{tabular}{C{2cm}C{2cm} p{10cm}}
\hline\hline
\textbf{Study}& \textbf{Domain} & \textbf{Definition} (elaborated or adopted)\\
\hline
\cite{Martin1995} & S\&T policy 	& "A 'generic emerging technology' is defined [...] as a technology the exploitation of which will yield benefits for a wide range of sectors of the economy and/or society" (p. 165)\\
\cite{Day2000}	& Management & 	"[...] emerging technologies as science-based innovation that have the potential to create a new industry or transform an existing ones. They include discontinuous innovations derived from radical innovations [...] as well as more evolutionary technologies formed by the convergence of previously separate research streams" (p. 30)\\
\cite{Porter2002} & S\&T policy & "Emerging technologies are defined [...] as those that could exert much enhanced economic influence in the coming (roughly) 15-year horizon." (p. 189)\\
\cite{Corrocher2003} & Evolutionary economics	& "The emergence of a new technology is conceptualised [...] as an evolutionary process of technical, institutional and social change, which occurs simultaneously at three levels: the level of individual firms or research laboratories, the level of social and institutional context, and the level of the nature and evolution of knowledge and the related technological regime." (p. 4) \\
\cite{Hung2006} & S\&T policy & "Emerging technologies are the core technologies, which have not yet demonstrated potential for changing the basis of competition" (p. 104)\\
\cite{Boon2008} & S\&T policy & "Emerging technologies are technologies in an early phase of development. This implies that several aspects, such as the characteristics of the technology and its context of use or the configuration of the actor network and their related roles are still uncertain and non-specific" (p. 1915)\\
\cite{Srinivasan2008} & Management & "I conceptualize emerging technologies in terms of three broad subheads: their sources [...], their characteristics [...] and their effects [...] Specifically, I consider two aspects of the sources of emerging technologies --- the 'relay race evolution' of emerging technologies, and 'revolution by application' --- four characteristics of emerging technologies --- the clockspeed nature of emerging technologies, convergence, dominant designs, and network effects --- and three effects of emerging technologies --- shifting value chains, digitization of goods, and the shifting locus of innovation (from within the firm to outside the firm)." (pp. 633-634)	\\
\cite{Cozzens2010}	& S\&T policy	& "Emerging technology --- a technology that shows high potential but hasn't demonstrated its value or settled down into any kind of consensus." (p. 364)
"The concepts reflected in the definitions of emerging technologies, however, can be summarised four-fold as follows: (1) fast recent growth; (2) in the process of transition and/or change; (3) market or economic potential that is not exploited fully yet; (4) increasingly science-based." (pp. 365-366) 
\\
\cite{Stahl2011} & S\&T policy & "[...] emerging technologies are defined as those technologies that have the potential to gain social relevance within the next 10 to 15 years. This means that they are currently at an early stage of their development process. At the same time, they have already moved beyond the purely conceptual stage. [...] Despite this, these emerging technologies are not yet clearly defined. Their exact forms, capabilities, constraints, and uses are still in flux" (pp. 3-4)\\
\cite{Alexander2012} & S\&T policy & "Technical emergence is the phase during which a concept or construct is adopted and iterated by [...] members of an expert community of practice, resulting in a fundamental change in (or significant extension of) human understanding or capability." (p. 1289)\\
\cite{Halaweh2013} & Management & Characteristics of (IT) emerging technologies "are uncertainty, network effect, unseen social and ethical concerns, cost, limitation to particular countries, and a lack of investigation and research." (p. 108)\\
\cite{Small2014} & Scientometrics & "[...] there is nearly universal agreement on two properties associated with emergence --- novelty (or newness) and growth." (p. 2)\\	
\hline\hline
\multicolumn{3}{l}{\footnotesize \textit{Source: search performed by authors on SCOPUS and extended to cited references.}}
\end{tabular}
}
\end{table}
%=====

The first defining attribute of emerging technology, explicitly included in two of the 12 core articles, is \textit{radical novelty}:  "novelty (or newness)" \citep{Small2014} may take the form of "discontinuous innovations derived from radical innovations" \citep{Day2000} and may appear either in the method or the function of the technology. To achieve a new or a changed purpose/function, emerging technologies build on different basic principles \citep{Arthur2007} (e.g.\ cars with an internal combustion engine vs.\ an electric engine, cytology-based techniques vs.\ molecular biology technologies).\ Novelty is not only a characteristic of technologies deriving from technical revolutions, i.e.\ technologies with relatively limited prior developments (e.g.\ DNA sequencing technologies, molecular biology, nano-materials), but it may also be generated by putting an existing technology to a new use. The evolutionary theory of technological change views this as the speciation process of technology, that is the process of applying an existing technology from one domain to another domain or 'niche' \citep{Adner2002}. The niche is characterised by a selection process that is different from the one where the technology was initially applied. The niche specifically may differ in terms of adaptation (the needs of the niche) and abundance of resources. The technology applied in the niche may adapt and then emerge as well as potentially invading other domains including the initial domain (giving rise to a 'revolution' or a process of 'creative destruction'). This implies that 'evolutionary' technology (those not characterised by revolutionary technical developments) can also be radically novel in domains of application different from those where the technology was initially developed. \cite{Adner2002} provided a compelling example of the speciation process by reporting on the evolution of wireless communication technology. This technology was created for laboratory purposes, and specifically for the measurement of electromagnetic waves. Yet, it found numerous subsequent applications. Wireless communication technology first enabled communication with locations (e.g.\ lighthouses) otherwise not reachable with wired telegraphy. Then, applications expanded to the transmission of voice (radiotelephony and broadcasting), and, more recently, to data transmission (Wi-Fi). With each shift, wireless communication technology appeared radically novel in its new domain of application, although the technology itself had existed since the early laboratory and telegraphy applications. The evolutionary theory of technological change teaches us that radical novelty may characterise innovations based on both revolutionary and evolutionary inventions resulting from the speciation process. However, the term 'evolutionary' is also used to refer to incremental technological advances. To avoid ambiguity, we opted to use the term 'radical novelty' rather than 'revolutionary/evolutionary' and to contextualise it in relation to the domain(s) in which the technology is arising.\footnote{The word 'novelty' alone may also create ambiguity with regard to the types of technologies we aim to include in our conceptualisation of emerging technologies. Technologies of a more incremental nature, as derived from the improvement of existing technologies, are somewhat novel. For the sake of conceptual clarity, we therefore prefer to add the attribute 'radical' to the word 'novelty'. } 

The second defining attribute of emerging technologies, identified by three of the 12 core articles is "clockspeed nature" \citep{Srinivasan2008} or "fast growth" \citep{Cozzens2010}, or at least "growth" \citep{Small2014}. Growth may be observed across a number of dimensions such as the number of actors involved (e.g.\ scientists, universities, firms, users), public and private funding, knowledge outputs produced (e.g.\ publications, patents), prototypes, products and services, etc. As with the radical novelty attribute, the fast growth of a technology needs to be contextualised. A technology may grow rapidly in comparison with other technologies in the same domain(s), therefore \textit{relatively fast growth} may be a better term.

The third attribute of emerging technologies, identified by four of the 12 core articles is \textit{coherence} that persists over time. The core articles variously describe this attribute as "convergence of previously separated research streams" \citep{Day2000}, "convergence in technologies" \citep{Srinivasan2008}, and technologies that "have already moved beyond the purely conceptual stage" \citep{Stahl2011}. \cite{Alexander2012} point instead to the role of "an expert community of practice", which adopts and iterates the concepts or constructs underlying the particular emerging technology. The concept of a community of practice suggests that both a number of people and a professional connection between those people are necessary. Coming together, intertwining and staying together are all entailed in coherence. Coherence refers to internal characteristics of a group such as 'sticking together', 'being united', 'logical interconnection' and 'congruity'. The status of external relations is also important. The emerging technology must detach itself from its technological 'parents' to some degree to merit a separate identity. Furthermore, it must stay detached for some period of time to be seen as self-sustaining \citep{Glanzel2012}. As we stated above, emergence is a process and coherence, detachment and identity do not characterise a final state, but are always in the process of realisation, presenting challenging issues of boundary delineation and classification. Perspective matters since an analyst may see an exciting emerging technology about to make a major economic impact in something a scientist sees as long past the exciting emerging phase.

The fourth defining attribute of emerging technologies, identified by nine of the 12 core articles is to yield "benefits for a wide range of sectors" \citep{Martin1995}, "create new industry or transform existing ones" \citep{Day2000}, "exert much enhanced economic influence" \citep{Porter2002}, or change "the basis of competition" \citep{Hung2006}. \cite{Corrocher2003} also point to the pervasiveness of the impact that the emerging technology may exert by crosscutting multiple levels of the socio-economic system, i.e.\ organisations and institutions, as well as knowledge production processes and technological regimes. Accordingly, we identify \textit{prominent impact} as another key attribute of emerging technologies. Most of the core articles conceived the prominent impact of emerging technologies as exerted on the entire socio-economic system. In this usage the concept of emerging technologies becomes very close to that of 'general purpose technologies' and so excludes technologies prominent within a specific domain. We wish to include relatively smaller scale prominence in our definition. For example, a diagnostic technology may emerge and significantly reshape the clinical practices associated with a given disease, profoundly affecting one disease domain but not others. In other words, our definition allows for prominent impact with narrow scope (emergence in one or a few domains), as well as wide-ranging impact across domains and potentially the entire socio-economic system (e.g.\ ICT and molecular biology). Such a perspective suggests, as with the attributes of radical novelty and relatively fast growth, the importance of contextualising the prominent impact of the observed technology within the domain(s) from which the technology emerges.

The final defining attribute of emerging technologies, identified in seven of the 12 core articles is that the prominent impact of emerging technologies lies somewhere in the future --- the technology is not finished. Thus, \textit{uncertainty} features in the emergence process. The non-linear and multi-factor nature of emergence provides emergence with a certain degree of autonomy, which in turn makes predicting a difficult task \citep{DeHaan2006,Mitchell2007}. As a consequence, knowledge of the probabilities associated with each possible outcome (e.g.\ potential applications of the technology, financial support for its development, standards, production costs) may be particularly problematic \citep{Stirling2007a}. Core articles expressed this attribute in terms of the 'potential' that emerging technologies have for changing the existing 'ways of doing things' \citep[e.g.][]{Boon2008,Hung2006,Stahl2011}. 

However, these definitions seem not to disentangle explicitly another important aspect of emergence from the concept of uncertainty. This is \textit{ambiguity}. Ambiguity arises because proposed applications are still malleable, fluid and in some cases contradictory, i.e.\ even the knowledge of possible outcomes of emergence is incomplete.  A variety of possible outcomes may occur because social groups encountered during emergence hold diverging values and ascribe different meanings to the technology \citep{Mitchell2007}. It is worth noting that uncertainty and ambiguity are, however, not mutually exclusive \citep{Stirling2007a}. These are not discrete conditions. A continuum exists as defined by the extent to which knowledge of possible outcomes and likelihood for each outcome is incomplete. For example, it may be problematic evaluating the probabilities associated with known possible outcomes, but at the same time there may also be a lack of knowledge of other possible outcomes such as unintended/undesirable consequences deriving from the (potentially uncontrolled) use of the technology. Uncertainty and ambiguity are key starting concepts for a wide variety of science and technology studies (STS) focusing on the role of the expectations in technological emergence \citep[e.g.][]{VanLente1998}.

The studies reviewed here introduced various additional concepts such as the science-based-ness, network effects, and early-stage development of emerging technologies. While the last of these seems to be implicit in the definition of emergence and the key role of networks (of users adopting the technology) is certainly not a unique feature of emerging technologies, the association with science-based-ness is less clear. The importance of science (especially public science) for the development of industrial technologies is widely accepted on the basis of substantial evidence \citep[e.g.][]{Narin1997}. However, even today not all technological revolutions may depend on breakthrough advances in science. In certain domains, a technology can be developed without the need for deep scientific understanding of how the phenomenon underlying it works --- "it is possible to know how to produce an effect without knowing how an effect is produced" \citep[p. 4]{Nightingale2014}. For example, \cite{Vincenti1984} provided evidence of this in the case of the construction of airplanes in the 1930s. The different parts of an airplane were initially joined using rivets with dome-shaped heads. These types of rivets, however, caused resistance to the air, thus reducing the aerodynamic efficiency of the plane. As other dimensions of airplane performance were improving (e.g.\ speed), the aerodynamic efficiency became increasingly relevant. The dome-shaped rivets were therefore replaced with rivets flush with the surface of the airplane. This was a major improvement for the aerodynamics of airplanes in 1930s, but it required no major scientific breakthrough.\footnote{Other classical examples include prehistoric cave dwellers using fire for cooking without any scientific understanding of it, the development of steam engines that predated the development of thermodynamics, or the Wright brothers testing flying devices before the field of aerodynamics was established.} A more recent example is the development of smartphones which did not require major advancements in science since most of the technologies used already existed --- the integration of these technologies, and advances in design for the creation of novel user interfaces instead provided the foundation of the innovation.\footnote{The innovation was architectural rather than modular according to the distinction proposed by \cite{Henderson1990}. Also, smartphone technology can be considered as an example of emerging technology of an evolutionary nature. As discussed above, the radical novelty of this technology is the result of existing technologies converging in new domains of applications.} For these reasons, 'science-based-ness' does not feature in our definition of emerging technologies.

In summary, as reported in \Tabref{match}, our review of innovation studies identified five main defining characteristics or attributes of emerging technologies: (i) radical novelty, (ii) relatively fast growth, (iii) coherence, (iv) prominent impact, and (v) uncertainty and ambiguity. Combining these attributes, we define an emerging technology as \textit{a radically novel and relatively fast growing technology characterised by a certain degree of coherence persisting over time and with the potential to exert a considerable impact on the socio-economic domain(s) which is observed in terms of the composition of actors, institutions and patterns of interactions among those, along with the associated knowledge production processes. Its most prominent impact, however, lies in the future and so in the emergence phase is still somewhat uncertain and ambiguous}.

%=====
\setlength{\tabcolsep}{10pt}
\renewcommand{\arraystretch}{1.5}
\begin{table}\footnotesize
	\caption{\label{tab:match}Attributes of emergence and reviewed key innovation studies.}
	\centering
{\begin{tabular}{lllllllllllll}
\hline\hline
  & \multicolumn{12}{c}{\textbf{Innovation studies defining emerging technologies}} \\
  \cline{2-13}
\textbf{Attribute of emergence} &   \rotatebox{90}{\scriptsize\cite{Martin1995}} &  \rotatebox{90}{\scriptsize\cite{Day2000}} & \rotatebox{90}{\scriptsize\cite{Porter2002}} & \rotatebox{90}{\scriptsize\cite{Corrocher2003}}  & \rotatebox{90}{\scriptsize\cite{Hung2006}}  & \rotatebox{90}{\scriptsize\cite{Boon2008}}  & \rotatebox{90}{\scriptsize\cite{Srinivasan2008}}  & \rotatebox{90}{\scriptsize\cite{Cozzens2010}} &
\rotatebox{90}{\scriptsize\cite{Stahl2011}} &
\rotatebox{90}{\scriptsize\cite{Alexander2012}} & 
\rotatebox{90}{\scriptsize\cite{Halaweh2013}} & 
\rotatebox{90}{\scriptsize\cite{Small2014}}\\
\hline
Radical novelty 			& 	&x	 & 	& 	&	 &	 & 	& 	& 	&	 &	&x\\
Relatively fast growth 		& 	&	 & 	& 	&	 &	 &x 	&x 	& 	&	 &	&x\\
Coherence				& 	&x	 & 	& 	&	 &	 &x 	&	&x	&x	 &	&\\
Prominent impact 			&x 	&x	 &x 	&x 	&x	 &	 &x 	&x 	&x	&x	 &	&\\
Uncertainty and ambiguity		& 	&x	 &x 	& 	&x	 &x	 & 	&x 	&x	&	&x	&\\
\hline\hline
\multicolumn{3}{l}{\footnotesize \textit{Source: authors' elaboration.}}
\end{tabular}
}
\end{table}
%=====

\begin{figure}
\includegraphics[width=13cm]{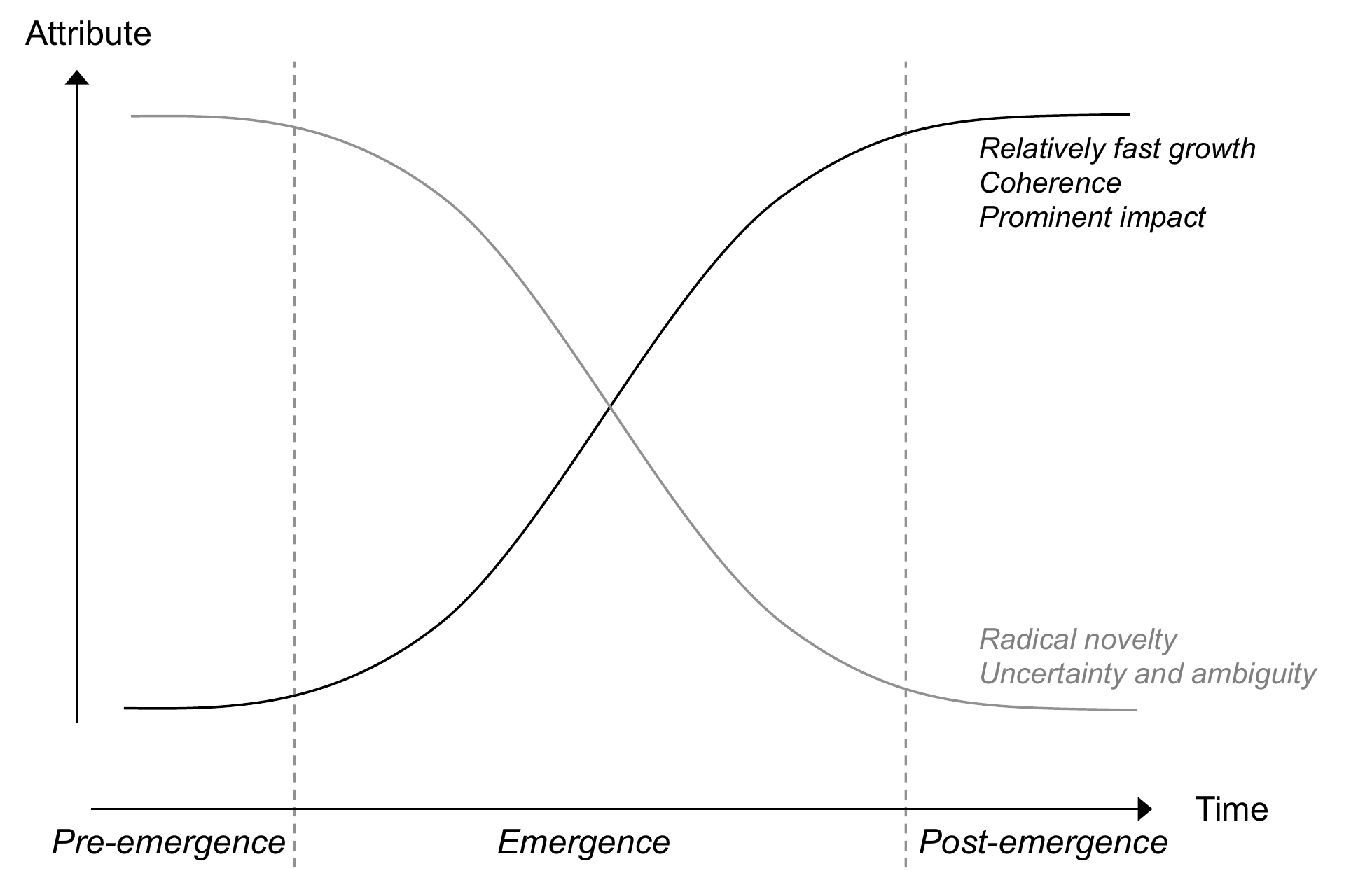}
\centering
\caption{Pre-emergence, emergence, and post-emergence: attributes and 'stylised' trends. \newline\textit{Source: authors' elaboration.}}
\label{fig:phases}
\end{figure}

It is reasonable to assume that the attributes of emergence range from 'low' to 'high' levels. Nonetheless, to try and pin them down to some absolute level is rather meaningless. As discussed, the attributes of emergence (especially radical novelty and relatively fast growth) provide an indication of emergence when they are considered in the domain in which the given technology is arising and therefore in relation to other technologies that may exist in that domain. Most importantly, these attributes are likely to co-evolve and assume very different levels over different periods of emergence. In the early stage of emergence ('pre-emergence'), a technology is likely to be characterised by high levels of radical novelty as compared to other technologies in the domain in which it is arising. However, the impact the technology can exert on that domain is still relatively low. The technology has not yet gone beyond the purely conceptual stage, multiple communities are involved in its development, and the delineation of the boundary of the technology is particularly problematic (i.e.\ low levels of coherence). As a consequence, its growth is relatively slow or not yet begun, and high levels of uncertainty and ambiguity are associated with the future developments of the technology --- the technology may not even emerge. The technology may then acquire a certain momentum. Some trajectories of development may have been selected out and certain dimensions of performance prioritised and improved. A community of practice may have also emerged. The technology thus becomes more coherent. Its impact is also relatively less uncertain and ambiguous, and the technology starts to take off in terms of publications, patents, researchers, firms, prototypes/products, etc. However, at the same time, it is likely that the radical novelty of the technology will diminish --- other technologies that exploit different basic principles may be emerging as well in the domain in which the considered technology is emerging. We conceived 'emergence' as this phase where the attributes of emergence are subject to dramatic change. Finally, impact and growth may enter a stable or declining phase, the technology loses its radical novelty, knowledge of the possible outcomes of the technology becomes more complete (probabilities can be perhaps assigned to outcomes), and the community of practice may become well-established (e.g. regular conferences, dedicated journals). The technology enters in a 'post-emergence' period. In line with the S-shaped patterns highlighted in early studies on the growth of science \citep[e.g][]{DeSollaPrice1963} and in technological adoption literature \citep[e.g.][]{Mansfield1961,Rogers1962}, we 'stylised' the change in the levels of the attributes of emergence as following an S curve (or more strictly, a reversed S curve in two of the five cases). This is qualitatively depicted in \Figref{phases}.  

Defining 'emerging technology' is, however, only half the battle.  If the definition is to be useful, we must show how the attributes can be measured and thus how technologies can be classified as emerging or not. In the next section, we  link our definition to the the operationalisation of our definition of emerging technologies.  We rely mainly on scientometric techniques, bringing in other approaches to fill certain gaps. 

%=====
%=====

%-------------------------------------------------------------------------------
%	A framework for the operationalisation of emergence
%-------------------------------------------------------------------------------
\section{A framework for the operationalisation of emergence}
Scientometric research has developed methods to detect emergence in science and technology and is therefore central to operationalising our definition. From the vast literature that touches on emerging technologies, we drew upon studies that offer ideas on operationalising our five attributes. We identified relevant scientometric studies by including the term 'topic' in the search string we used to select research works dealing with definitional issues of emerging technologies ---  'topic' is often used in scientometrics to refer to the emergence of a new set of research activities in science and technology \citep[e.g.][]{Small2014,Glanzel2012}. The search was also extended to publication titles, abstracts and keywords, but narrowed to journals mainly or to a significant extent oriented toward the publication of novel scientometric techniques (see the right-hand column of \Tabref{searches}). The search in SCOPUS returned 155 publications. 

The examination of cited references of these publications enabled us to retrieve additional studies that were not captured with the search string, but are potentially relevant to for our analysis. This increased the initial sample to 183 studies. We then analysed these publications to identify studies that were relevant to the operationalisation of the attributes of emergence. This process led to a final set of 55 publications,\footnote{We excluded 76 studies that did not operationalise emergence (e.g.\ use of emerging technologies as empirical context for various analyses, examination of ethical issues associated with emerging technologies), three studies focused on the review of scientometric methods for the analysis of emerging technologies, and two studies elaborating document search strategies based on a modular lexical approach. 33 studies that were concerned with Future-oriented Technology Analysis (FTA) techniques (e.g.\ foresight, forecasting, roadmapping, Constructive Technology Assessment (CTA)) were also not included in the review. While about 67\% of these do not rely on scientometrics, the remaining FTA studies in the sample propose frameworks for selecting, rather than identifying, emerging technologies, or adopt conventional scientometric/bibliometric approaches, which will be instead discussed with the review of the selected scientometric studies. FTA methods, however, remain crucial for more prospective analyses of emerging technologies and decision-making on possible future scenarios \citep[e.g.][]{Porter2004, Irvine1984, Ciarli2013}. 14 STS studies included in the sample will be instead referenced in our review and discussion when the operationalisation of the attributes of emergence with the use of scientometric approaches is limited by a lack of data or by the nature of the considered attribute. It is worth noting that our search did not capture 'technometric' studies \citep[e.g.][]{Grupp1994,Saviotti1984,Sahal1985}. This research stream has been particularly important for the measurement of technology and technological change. Nonetheless, technometric models tend to rely on a variety of assumptions and often require data, the collection of which can be particularly labour-intensive (e.g.\ extraction and coding of data on the features of the considered technologies) \citep[e.g.][]{Coccia2005}.} which were then classified in terms of the methodological approach adopted to detect or analyse emergence (e.g.\ indicators, citation patterns between documents, co-occurrence of words in text), data sources used (e.g.\ publications, patents, news articles), and proposed operationalisation of emergence. This information is summarised in \Tabref{methods} where studies are grouped into five groups: (i) indicators and trend analysis studies that are mainly based on document counts; (ii) citation analysis studies which focus on examining citation patterns between documents; (iii) co-word analysis studies that build on the co-occurrence of words across document text; (iv) overlay mapping technique studies, which use projections to position a given set of documents within a wider or more global structure (e.g.\ a map of science); and (v) hybrid studies that combine two or more of the above approaches. \Tabref{methods} shows how definitions of emergence varied, even within the same group of techniques, thus providing further evidence of the low level of consensus on what constitutes emergence. 

Given the definitional weaknesses in the original studies, our use of a particular study often varies from that of its authors. We will briefly introduce the major techniques and our interpretation of the contribution they make to measuring attributes of emerging technologies. For each attribute, we will first describe how it can be operationalised for contemporary and then for retrospective cases of emerging technologies. When data scarcity or the nature of the attribute of emergence limit the applicability of scientometrics, we will discuss qualitative approaches. The role of experts remains crucial for the validation of the results obtained with the use of the techniques discussed below, especially for qualitative approaches to the operationalisation of emergence.

%=====
%====
\afterpage{
\begin{landscape}\scriptsize
\begin{spacing}{0.8}
\begin{longtable}{p{6cm}p{2.7cm}p{13cm}}
\caption{\label{tab:methods}Methods for the detection and analysis of emergence in science and technology (studies are ordered by technique and publication year).} \\
\endfirsthead
\caption{Methods for the detection and analysis of emergence in science and technology (studies are ordered by technique and publication year) \textit{(continued)}.} \\
\hline\hline
\endhead
\hline\hline
\textbf{Method/Study}&			\textbf{Data}&			\textbf{Operationalisation of emergence}\\
\hline
\textbf{Indicators and trends}\\
\addtolength{\leftskip}{1em}\noindent		\cite{Porter1995}&  			Publications/patents 		& Count of keywords in publication abstracts and trend analysis based on Fisher-Pry curves \\ [1ex]
\addtolength{\leftskip}{1em}\noindent		\cite{Kleinberg2002}&  		Publications/e-mails		& 'Burst of activity' detected as state transitions of an infinite-state automaton\\	[1ex]
\addtolength{\leftskip}{1em}\noindent		\cite{Bengisu2003}&  		Publications			& Positive slope of the line derived by regressing the number of publications on time and no decrease of more than 10\% or stability (no increase) in the last period or continuos decline in the last three periods of observation\\ [1ex]	
\addtolength{\leftskip}{1em}\noindent		\cite{Watts2003}&  			Publications			& Indicators of emergence: cohesion (based on cosine similarity between documents), entropy, and F-measure \\ [1ex] 
\addtolength{\leftskip}{1em}\noindent		\cite{Bettencourt2008}&  		Publications			& Epidemic model to describe the increasing number of authors involved in an emerging field \\ [1ex] 
\addtolength{\leftskip}{1em}\noindent		\cite{Bettencourt2009}&  		Publications			& Increasing densification (average number of edges per node), stable/decreasing diameter (average path length between nodes), and increasing fractional count of edges in the largest component of the co-authorship network\\ [1ex] 
\addtolength{\leftskip}{1em}\noindent		\cite{Moed2010}&  		Publications		& Journals characterised by high values of Source Normalised Impact per Paper (SNIP) indicator\\	[1ex]
\addtolength{\leftskip}{1em}\noindent		\cite{Schiebel2010,Roche2010} &  Publications			& Publication keywords initially labelled as "unusual terms", by using \textit{tf-idf} and Gini coefficient, that subsequently become "cross section terms", i.e. they diffuse in several research domains\\ [1ex]
\addtolength{\leftskip}{1em}\noindent		\cite{Guo2011}&	  		Publications			& Indicators of emergence: frequency of keywords (ISI keywords, authors' keywords, and MeSH terms), growing number of authors, and interdisciplinarity (based year-average Rao-Stirling diversity index) of cited references\\ [1ex]
\addtolength{\leftskip}{1em}\noindent		\cite{Jarvenpaa2011}&  		Mixed				& Absolute and cumulative count of the number of basic and applied research publications, patents, and news\\ [1ex]
\addtolength{\leftskip}{1em}\noindent		\cite{Abercrombie2012}&  		Mixed				&Normalised number of publications and citations, patents, and web news fitted to a polynomial function\\ [1ex]
\addtolength{\leftskip}{1em}\noindent		\cite{Jun2012,Jun2014}&  			News			& Normalised searching traffic (Google trends)\\ [1ex]
\addtolength{\leftskip}{1em}\noindent		\cite{Avila-Robinson2013a,Avila-Robinson2013}&  Publications/patents &  Overview of indicators to analyse emergence\\ [1ex]
\addtolength{\leftskip}{1em}\noindent		\cite{DeRassenfosse2013}&  	Patents				& Count of the priority patent applications filed by a country's inventor, regardless of the patent office in which the application is filed \\ [1ex]
\addtolength{\leftskip}{1em}\noindent		\cite{Ho2014}&  			Publications			& Cumulative number of publications fitted to a logistic curve\\ [1ex]
\hline

\newpage
\textbf{Method/Study}&			\textbf{Data}&			\textbf{Operationalisation of emergence}\\
\hline
\textbf{Citations analysis}\\
\textit{Direct citation}\\
\addtolength{\leftskip}{1em}\noindent		\textit{Seminal paper:} \cite{Garfield1964}&  	Publications&   - \\ [1ex]
\addtolength{\leftskip}{1em}\noindent		\cite{Kajikawa2008,Kajikawa2008a,Takeda2008}&  		Publications&  	Clusters of publications with the highest average publication year\\ [1ex]
\addtolength{\leftskip}{1em}\noindent		\cite{Scharnhorst2010}&  		Publications&   Historiographic approach combined with 'field mobility' of publications\\ [1ex]
\addtolength{\leftskip}{1em}\noindent		\cite{Shibata2011}&  			Publications&  	Clusters of publications with the highest values of betweenness centrality\\ [1ex]
\addtolength{\leftskip}{1em}\noindent		\cite{Iwami2014}&  		Publications&  	Publications ('leading papers') with high values of in-degree ('height'), large variation of in-degree between one year and the next year ('slope'), or large cumulative in-degree ('area') as defined on the basis of the yearly direct citation network\\ [1ex]

\textit{Co-citation}\\
\addtolength{\leftskip}{1em}\noindent		\textit{Seminal paper:} \cite{Small1973}&  	Publications&  		-\\ [1ex]
\addtolength{\leftskip}{1em}\noindent		\cite{Small2006}&			Publications&			Clusters with no continuing publications from the prior period\\ [1ex]
\addtolength{\leftskip}{1em}\noindent		\cite{Cho2011}&  			Patents&  				Technological patent classes (IPC) that span structural holes in the co-citation network\\ [1ex]
\addtolength{\leftskip}{1em}\noindent		\cite{Erdi2012}& 			Patents&  				Clusters of patents present in a given time period and not in the previous period\\ [1ex]
\addtolength{\leftskip}{1em}\noindent		\cite{Boyack2014}& 			Publications&  			Yearly clustered publications of which references overlap less than 30\%  with references cited by previous clusters\\ [1ex]

\textit{Bibliographic coupling}\\
\addtolength{\leftskip}{1em}\noindent		\textit{Seminal paper:} \cite{Kessler1963}&  Publications&  	-\\[1ex]
\addtolength{\leftskip}{1em}\noindent		\cite{Morris2003}&	Publications&	Clusters of publications that cite more recent clusters of publications, namely emerging research fronts \\[1ex]
\addtolength{\leftskip}{1em}\noindent		\cite{Kuusi2007}&	Patents&		Clusters of patents as source to identify guiding images ('leitbild') of technological development\\[1ex]
\hline

\textbf{Co-word analysis}\\
\addtolength{\leftskip}{1em}\noindent		\textit{Seminal paper:} \cite{Callon1983}&  		Publications&  		-\\[1ex]
\addtolength{\leftskip}{1em}\noindent		\cite{Lee2008}&  		Publications&  			Clusters in the co-word network that show low values of degree, high betweenness, and low closeness, i.e. those clusters that are more likely to turn into hub in the future. \\[1ex]
\addtolength{\leftskip}{1em}\noindent		\cite{Ohniwa2010}&  		Publications&  		 MeSH terms (clustered with co-word analysis) that are included in the top-5\% by incremental rate in a given year --- the increment rate for a MeSH term is defined as the number of time the terms occurred at the time $t$, $t+1$, and $t+2$ out the number of times the term occurred at $t-1$, $t$, $t+1$, and $t+2$\\[1ex]
\addtolength{\leftskip}{1em}\noindent		\cite{Yoon2011}&  		Patents&  			Small and dense sub-networks in the 'invention property-function' network\\[1ex]
\addtolength{\leftskip}{1em}\noindent		\cite{Furukawa2015}&  	Publications&  		Sessions of conferences in which previous sessions converge according to the average cosine similarity (based on \textit{tf-idf}-identified keywords) between the papers included in the sessions\\[1ex]
\addtolength{\leftskip}{1em}\noindent		\cite{Zhang2014}&  		Publications&  		Combination of cluster analysis with term clumping and principal component analysis\\[1ex]

\hline

\textbf{Method/Study}&			\textbf{Data}&			\textbf{Operationalisation of emergence}\\
\hline
\textbf{Overlay mapping}&  			&  \\
\addtolength{\leftskip}{1em}\noindent		\cite{Rafols2010}&  			Publications&  		Overlays of publications projected on a basemap of ISI WoS subject categories linked by cosine similarity of co-citations patterns between journals\\[1ex]
\addtolength{\leftskip}{1em}\noindent		\cite{Bornmann2011}&  		Publications&  		Overlays of publications on Google maps to identify cities publishing more than expected \\[1ex]
\addtolength{\leftskip}{1em}\noindent		\cite{Leydesdorff2011}& 		Publications& 		Overlays of publications and co-authorship networks on Google maps to trace collaboration activity \\[1ex]
\addtolength{\leftskip}{1em}\noindent		\cite{Leydesdorff2012}&  		Publications&  		Overlays of publications projected on a basemap of MeSH terms linked by cosine similarity (based on the co-occurrence of MeSH terms at the publication level) \\[1ex]
\addtolength{\leftskip}{1em}\noindent		\cite{Leydesdorff2012a}&  	Patents&  			Overlays of patents on Google maps to identify cities patenting more than expected \\[1ex]
\addtolength{\leftskip}{1em}\noindent		\cite{Leydesdorff2013j}&  		Publications&  		Overlays of publications projected on the basemap of journals linked by cosine similarity of co-citations patterns between journals\\[1ex]
\addtolength{\leftskip}{1em}\noindent		\cite{Kay2014}&  			Patents&  			Overlays of patents projected on the basemap of 466 IPC classes linked by cosine similarity of citing-to-cited relationships between classes --- the basemap is built by using patents included in 2011 PATSTAT\\[1ex]
\addtolength{\leftskip}{1em}\noindent		\cite{Leydesdorff2014}&  		Patents&  			Overlays of patents projected on the basemap of 124 3-digit or 630 4-digit IPC classes linked by  cosine similarity based on co-citations between classes --- the basemap is built by using patents granted at the United States Patent and Trademark Office (USPTO) from 1976 to 2011 \\[1ex]

\hline
\textbf{Hybrid}\\

\addtolength{\leftskip}{1em}\noindent		\cite{Chen2006}: co-citation analysis and burst detection&  		Publications&  		Trends in the bipartite network of research-front terms (burst detection) and intellectual base articles --- the network includes three types of links: co-occurring research front terms, co-cited intellectual base articles, and a research-front term citing an intellectual base article \\[1ex]

\addtolength{\leftskip}{1em}\noindent		\cite{Leydesdorff1994}: co-citation analysis and bibliographic coupling& 	Publications&		New journals that build on multiple existing areas, i.e. they load on multiple factors obtained by the factor analysis of the matrix of the cited references, and have unique 'being cited' patterns, i.e. they are 'central tendency journals' reporting highest load on a given factor as obtained by the factor-analysis of the matrix of received citations\\[1ex]

\addtolength{\leftskip}{1em}\noindent		\cite{Glanzel2012}: co-word, direct citation analyses and bibliographic coupling&  	Publications&  		Existing clusters with exceptional growth, completely new clusters with roots in other clusters, and existing clusters with a topic shift\\[1ex]

\addtolength{\leftskip}{1em}\noindent		 \cite{Gustafsson2015}: co-occurrence of IPC classes& Patents&		Technological co-classification to identify clusters of patents and detect guiding images or 'leitbild' from patent full-text\\[1ex]

\addtolength{\leftskip}{1em}\noindent		 \cite{Small2014}: direct and co-citation analyses&  		Publications&  		Clusters of publications that show high growth and are new both to the direct citation and co-citation models \\[1ex]

\addtolength{\leftskip}{1em}\noindent		\cite{Yan2014}: co-word analysis and topic modelling &  		Publications&  		Topics that are not a close variation of other topics, i.e. a topic $i$ in the year $t$ is emerging if no predecessors are found and no other topics are transformed into topic $i$ at $t+1$\\[1ex]

\addtolength{\leftskip}{1em}\noindent		\cite{Chang2009,Breitzman2015}: direct citation and co-citation analyses &  		Patents&  		Clusters of patents (co-citation clustering) that form around 'hot' patents --- defined as those patents that are highly cited (top 5\%-10\%) by patents issued in the last two years and the citations of which mostly come from patents issued in the last two years\\[1ex]

\hline\hline
\multicolumn{3}{l}{\footnotesize \textit{Source: search performed by authors on SCOPUS and extended to publication cited references.}}

\end{longtable}
\end{spacing}
\end{landscape}
}
%=====
%=====

%-------------------------------------------------------------------------------
\subsection{Radical novelty}
Emerging technologies are radically novel, i.e. they fulfill a given function by using a different basic principle as compared to what was used before to achieve a similar purpose. Publications and patents are of limited use in assessing radical novelty in contemporary technology. In contrast, news articles, editorials, review and perspective articles in professional as well as academic journals represent valuable sources, providing participant perspectives on if and why a technology is viewed as radically novel. These documents may also provide an understanding of the basic principles underpinning the examined technology. 

In contrast, in retrospective analyses citation and co-word analyses can be particularly effective for identifying radical novelty. Relatively large amounts of data can be exploited to map the cognitive networks of a knowledge domain over time. Citation analysis builds on citation patterns among documents to generate a network in which nodes are documents and links between nodes represent (i) a direct citation between two documents (direct citation analysis) \citep{Garfield1964}, (ii) the extent to which two documents are cited by the same documents (co-citation analysis) \citep{Small1973}, or (iii) to what extent two documents cite the same set of documents (bibliographic coupling) \citep{Kessler1963}. Co-word analysis instead exploits the text of documents to create a network of keywords (or key phrases) that are linked according to the text to which they co-occur across the set of selected documents \citep{Callon1983}. 

On the premise that clusters of documents or words in these networks represent different knowledge areas of a domain or different literatures on which the domain builds, several studies have considered the appearance of clusters not previously present in the network as a signal of novelty \citep[e.g.][]{Erdi2012,Kajikawa2008}. Others dispute this interpretation. Given the continuous evolution of science and technology, one is unlikely to find a cluster again in subsequent annual networks so the percentage of clusters that would qualify as newly appearing tends to be relatively high. For this reason, additional criteria have been suggested such as the appearance of new clusters that also link otherwise weakly connected (e.g.\ betweenness centrality) clusters \citep[e.g.][]{Shibata2011,Furukawa2015}, that form around documents that are highly cited by recent documents and the citations of which also are mostly from recent documents \citep{Breitzman2015}, or that cite more recent clusters as identified by the (Salton) similarity of their references \citep{Morris2003}.

\cite{Small2014} have recently proposed a hybrid approach based on a combination of direct citation and co-citation models as applied to publication data. This approach is particularly focused on the detection of novelty, which is defined in terms of clusters that are new to the co-citation model --- that is, clusters with limited overlap with the cited documents included in clusters in previous years \citep{Boyack2014} --- as well as to a parallel direct citation model. By combining bibliographic coupling, co-word analysis, and direct citation analysis, \cite{Glanzel2012} instead defined novelty (namely emerging topics) as three cases of clusters: those that show exceptional growth, those that are completely new but with their roots in other clusters, or already existing ones that exhibit a topic shift.  \cite{Yan2014} combined co-word analysis with Natural Language Process (NLP) approaches (topic modelling). Emergence, as reflected in novelty, is then associated with the appearance of topics that are not a close variation of other topics calculated on the basis of the Jenson-Shannon Divergence.\footnote{The Jenson-Shannon Divergence is a measure of similarity between empirically-determined distributions (e.g.\ co-occurrence of words in documents) based on Shannon entropy measures \citep[for more details see][]{Lin1991}.} Specifically, a topic $i$ appearing at time $t$ is considered to be emerging if it has no predecessors and none of the identified topics transforms into topic $i$ at $t+1$. A different perspective is provided by \cite{Scharnhorst2010} who extended the analysis of historiographs (based on direct citations) to trace the extent to which publications move across fields as they receive citations from new fields (namely 'field mobility'). Assuming that these publications are associated with a basic principle used for technological applications, this approach enables one to identify which fields may be using a different knowledge base and thus in which fields radically novel technologies are potentially emerging. However, this requires \textit{a priori} knowledge of the basic principle and the set of documents associated with it.

Research in scientometrics has also focused on the development of techniques to expand the 'local' (domain) perspective that citation or text-based approaches may provide. This effort has generated a number of overlay mapping techniques \citep[for an overview see][]{Rotolo2014c}, which in turn may be particularly well suited to detecting radical novelty. The basic idea is to project a given set of documents (e.g.\ publications associated with a research domain) on a \textit{basemap} through the use of an \textit{overlay}. The basemap can represent the 'global' science structure at the level of the scientific discipline (ISI Web of Science (WoS) subject categories) \citep[e.g.][]{Rafols2010}, journal \citep[e.g.][]{Leydesdorff2013j}, Medical Subject Headings (MeSH)  \citep{Leydesdorff2012}, or the technological structure at the level of patent classes \citep[e.g.][]{Kay2014,Leydesdorff2014}.\footnote{The elements of the basemap are linked according to similarity based on the co-occurrence of citations or, in the case of MeSH, the co-occurrence of terms. The same approach can be used to project a sample of publications and patents onto geographical maps (e.g.\ Google maps) to reveal the most active cities and collaborative activities (see\Tabref{methods}).}  Once the set of documents (publications or patents) associated with a given domain has been identified, the projection of these documents over different time slices on the global map of science or technology may reveal the increasing involvement of new scientific or technological areas. This may suggest that new knowledge areas are being accessed to conduct research, and thus that potentially different basic principles are drawn upon to achieve a given purpose.

Among the studies within the 'indicators and trends' group of techniques, \cite{Moed2010} proposed the source normalised impact per paper (SNIP) indicator for the evaluation of journals' impact and claims it is relevant for identifying emerging technologies. This indicator is defined as the ratio between the journal's \textit{raw impact per paper} (number of citations in the year of analysis to the journal's papers published in the three previous years, divided by the number of the journal's papers in these three years) and the \textit{relative database citation potential} in the subject field covered by the journal (mean number of 1-3-year-old references per paper citing the journal and published in journals included in the considered database divided by that for the median journal in the database). \cite{Moed2010} argued that the SNIP indicator, and specifically high values of this indicator, also provides information on the extent to which a considered journal covers emerging topics. Given the focus on recent citations and database coverage, the SNIP indicator is clearly associated with the radical novelty attribute of emergence. This indicator is, however, evaluated at the aggregate level of the journal and journal-by-journal. It is therefore less clear whether signals of radical novelty (i.e.\ relatively high values of SNIP) are associated with one or multiple emerging topics the considered journal may cover. In addition, the SNIP may not capture signals of radical novelty in those instances of journals that cover few emerging topics and therefore characterised by low values of SNIP.  

All these techniques have various advantages and limitations. The qualitative analysis of news articles, editorials, review and perspective articles, for example, may be effective for contemporary analyses.  However, the technical language used in these documents may be an important barrier to a non-expert's efforts to independently assess radical novelty. The application of citation and co-word analyses is strongly dependent on time. Data need to be longitudinal in order to permit the tracing of cognitive dynamics and associated changes in the knowledge structure. Co-word analysis and bibliographic coupling are, however, less sensitive to time than direct citation and co-citation analyses and can be applied as documents become available \citep[e.g.][]{Breitzman2015}. Finally, overlay mapping provides a global perspective on emergence for the assessment of radical novelty, but interpretation of the resulting maps is mainly based on visual inspection.

%-------------------------------------------------------------------------------
\subsection{Relatively fast growth}
Emerging technologies show relatively fast growth rates compared to non-emerging technologies. The assessment of this attribute is particularly problematic for contemporary analyses. Growth is not yet observed in terms of publications and patents, for example, so scientometric indicators cannot be used. Early indications of growth may be revealed from the analysis of funding data, big data, and altmetrics. This is an important research direction for future studies on the operationalisation of the relatively fast growth attribute, as we will discuss later in the paper.

 In the case of retrospective analyses, 'relatively fast growth' is perhaps the most frequently measured attribute of emergence in scientometrics. Most studies assume rapid growth as a \textit{sine qua non} condition of emergence, and so a number of operationalisation approaches have been proposed. Indicators and trend analyses based on the yearly or cumulative count of documents --- publications, patents or news articles, according to the nature of the examined technology and the availability of data --- over a given observation period are widely used. Documents are generally identified over time by using expert-defined keywords appearing in the publication titles and abstracts \citep[e.g.][]{Porter1995} or by exploiting more institutionalised vocabularies such as the MeSH classification in the case of publication counts in the biomedical domain \citep[e.g.][]{Guo2011}. With a focus on patent data, \cite{DeRassenfosse2013} proposed counting the priority patent applications filed by a country's inventor, regardless of the patent office in which the application is filed, as an indicator to identify fast growth and therefore potential emerging technologies. However, yearly publication or patent counts are always dynamic, so the problem becomes one of setting a criterion by which to distinguish the signal from the noise, that is differentiating emerging technology from other increasing trends. Some theoretical foundations are needed to do this. 

Rapid growth is also detected by fitting the document count to a function (e.g.\ forms of logistic function such as Fisher-Pry curves).\footnote{Fisher-Pry curves were developed to model technological substitution between two competing technologies \citep{Fisher1971}. This family of curves is built on the basis of three assumptions: (i) technological advancements are the results of competitive substitutions of one method (technology) used to satisfy a given need for another; (ii) the new technology completely replaces the old technology; and (iii) the market share follows Pearl's Law, i.e. "the fractional rate of fractional substitution of new for old is proportional to the remaining amount of the old left to be substituted" \citep[p. 75]{Fisher1971}.} \cite{Bengisu2003}, for example, regressed the number of publications over publication year and defined emerging technologies as those technologies showing a positive slope and a decrease of less than 10\% or stability (no increase) in the last period compared to the previous one, or no continuous decline in the last three periods of observation. \cite{Ho2014} instead fitted the cumulative number of publications to a logistic curve, whereas \cite{Abercrombie2012} extended the count of publications to patents, web news, and commercial applications. Data were then normalised and fitted to a polynomial function for comparison --- a similar approach is employed by \cite{Jarvenpaa2011} and \cite{Jun2014}. 

The number of documents is also used to detect 'bursts of activity', i.e.\ the appearance of a topic in a document stream. This relies on the approach of \cite{Kleinberg2002}, who modelled the number of publications and e-mails containing a given set of keywords as an infinite-state automaton, i.e.\ a self-operating virtual machine that may assume a non-finite number of states and where the transition from one state to another is regulated by a 'transition function' (similarly to Markov models). The frequency of state transitions with certain features identifies bursts of activity, which are used as a proxy for fast growth. The burst detection approach is combined with co-citation analysis by \cite{Chen2006} to build a bipartite network\footnote{A bipartite network is a network in which nodes can be partitioned into two distinct groups, $N_1$ and $N_2$, and all the links connect one node from $N_1$ with a node from $N_2$, or vice versa \citep{Wassermann1994}.} of research-fronts linked with intellectual base articles. This network is then analysed in order to identify emerging trends.

\cite{Schiebel2010} and \cite{Roche2010} proposed instead an approach to emergence that is based on a diffusion model (and diachronic cluster analysis to identify topics) that combines a modified \textit{tf-idf}\footnote{The \textit{tf-idf} (term frequency-inverse document frequency) is an indicator that reflects the importance of a word to a document in relation to a corpus. Specifically, the \textit{tf-idf} is the result of the product between two indicators: the term frequency and inverse document frequency.} with the Gini coefficient to identify three stages:  "unusual terms", "established terms", and "cross section terms". Unusual terms are those that are rare in publications since they describe a research discovery at a very early stage. When research intensifies, terms first become more established in the original domain and subsequently they may diffuse into other domains, thus becoming cross section terms. Terms that change their classification (i.e.\ that show pathways) from unusual to cross section terms from one period to another are characterised by rapid diffusion and therefore relatively fast growth. This approach, however, is highly dependent on the thresholds of the \textit{tf-idf} and Gini coefficient selected to classify terms as well as on the duration of the periods used to trace changes in the classification of terms.

Citation and co-word analyses can also be used to identify the relatively rapid growth of a potential emerging technology. Longitudinal analysis of the size of the clusters of documents or words obtained with the application of these techniques can detect knowledge areas that show rapid growth. For example, \cite{Ohniwa2010} used co-word analysis to cluster MeSH terms. For each MeSH term an increment rate was calculated in year $t$ as the number of times the term occurred at time $t+1$ and $t+2$ out of the number of times the term occurred at $t-1$, $t$, $t+1$, and $t+2$. Fast growing topics are those in the top 5\% of the increment rate in a given year. 

\cite{Glanzel2012} combined bibliographic coupling, co-word analysis, and a direct citation model. First, documents are clustered in time slices according to their cosine similarity resulting from bibliographic coupling and textual similarity. The core clusters identified through this process are next linked across different time slices via direct citations. Emergence is then detected by identifying clusters with exceptional growth --- the study also considers emerging clusters to be those that are completely new with roots in other clusters or existing clusters exhibiting a topic shift, but this clearly refers to the radical novelty attribute of emergence. 

Overlay mapping techniques can visually reveal knowledge areas characterised by a rapid increase in the number of documents (publications or patents) in the 'global' maps of science or technology and which therefore, in comparison with other areas, may be growing at a faster pace. (Overlay mapping can also reveal diffusion across disciplines and technological areas.)

Other studies instead operationalised relatively fast growth by examining the growing number of authors involved in an emerging field over time \citep[e.g.][]{Guo2011,Bettencourt2008}. For example, \citep{Bettencourt2008} found that the growth of the population of authors in a given field tends to be relatively well described with epidemic models that consider novel ideas as spreading by 'infecting' authors.

%-------------------------------------------------------------------------------
\subsection{Coherence}
Coherence and its persistence over time distinguish technologies that have acquired a certain identity and momentum from those still in a state of flux and therefore not yet emerging. Coherence in contemporary technologies may be detected by examining the scientific discourse around a given emerging technology. Initially, a variety of terms may be in use and reduction in the number of terms may signal increasing coherence. Abbreviations or acronyms take time to appear and, when they do, signal persistence; they also indicate shared interpretations and thus coherence \citep{Reardon2014}. Additional signals of coherence may come from the creation of conference sessions, tracks, dedicated conferences and subsequently from journal special issues and new specialist journals \citep{Leydesdorff1994}. New categories in established classification systems may also be created \citep{Cozzens2010}.

In retrospective analyses, entropy measures can be used \citep{Watts2003} as well as clustering and factor analysis of citation and text networks. The coherence of clusters of documents or terms can be assessed in comparison to the overall network by applying, for example, local network density measures as well as by examining cluster persistence over time. \cite{Furukawa2015} propose using year-to-year coherence of conference sessions to indicate emergence. They applied co-word analysis to generate 'chronological' networks of conference sessions (nodes) linked by their (cosine) similarity as based on the keywords included in the sessions' papers --- keywords were selected using the \textit{tf-idf} indicator. Within these networks, emerging topics are defined as sessions where previous conferences' sessions converge according to similarity. 

In a similar vein, \cite{Yoon2011} developed a NLP algorithm capable of identifying properties and functions in the sentences of patent abstracts.\footnote{This enables one to overcome the main limitation of co-word analysis techniques, that is the need to define an initial set of keywords before the analysis can be performed.} The method generates an 'invention property-function' network (IPFN). Nodes in this network represent properties and functions. A property is \textit{what a system is or has} and is expressed by using 'adjectives+nouns', whereas a function is \textit{what a system does} and is expressed by using 'verbs+nouns'. Links between nodes are defined by the co-occurrence of properties and functions in patents. Emerging properties and functions are those clustered in small and highly dense sub-networks --- i.e.\ \textit{de facto} showing a certain degree of coherence.

The approaches discussed above examine cognitive dynamics. However, coherence can also be assessed on the basis of changes in the social structure. In this regard, \cite{Bettencourt2009} examined the evolution of co-authorship networks at the level of the scientist to identify network patterns associated with the emergence of new scientific fields. Increasing average number of edges per nodes (densification), stable or decreasing average path length between two nodes (diameter), and increasing fractional count of edges in the largest component of the considered network were suggested as signals of emergence and specifically of the topical transition of a field. These indicators clearly refer to increasing connectedness of the co-authorship network, identifying emerging communities as an indicator of emerging technology.

%-------------------------------------------------------------------------------
\subsection{Prominent impact}
Emerging technologies exert a prominent impact on specific domains or more broadly on the socio-economic system by changing the composition of actors, institutions, patterns of interactions among those, and the associated knowledge production processes. Scientometric methods cannot identify contemporary prominent impact due to a lack of data and the difficulty in delineating the technology in its very early stages (e.g. keywords may still be used by groups of actors with different meanings and in different contexts). Mixed qualitative-quantitative approaches used by Science and Technology Studies (STS) scholars on the role of expectations in driving technological change are of a particular relevance.\footnote{Scientometrics can be considered as the more quantitative end of STS work. For this reason, the distinction we make between the two traditions is not intended to be a particularly strong one. However, it also true that there has been relatively little interaction between scientometrics and STS since the late1980s. Each of these tradition has its own conferences and journals, and only a handful of researchers operate at the interface --- most individuals would identify themselves as either 'scientometricians' or 'STS' scholars.} The main argument of the sociology of expectations is that "novel technologies and fundamental changes in scientific principle do not substantively pre-exist themselves, except and only in terms of the imaginings, expectations and visions that have shaped their potential" \citep[p. 285]{Borup2006}. These expectations are "real-time representations of future technological situations and capabilities [...] wishful enactments of a desired future" \citep[p. 285]{Borup2006} and they play a generative role by stimulating and steering as well as coordinating actions. Evidence of this has been found in a number of emerging fields such as gene therapy, pharmacogenomics, and nanotechnology \citep[e.g.][]{Selin2007,Hedgecoe2003,Martin1999}.  Expectations of the performance of novel technologies or, more generally, the ability of novel technologies to address societal problems are both important. 

News articles, editorials, review and perspective articles in professional and academic journals, vision reports and technological roadmaps have all been used to identify statements representing multiple and potentially competing expectations surrounding a technology \citep[e.g.][]{Alkemade2012,VanLente2010,Bakker2011}.  STS work has also illuminated the central role played by hype in technology emergence. Actors who understand the constitutive role of expectations have an incentive to raise expectations in order to motivate the funding and activity needed to realise their preferred technological future. Hype, or over-claimed expectation, is often the result.  This over-claiming can touch most attributes of emergence and especially prominent impact. For example, press releases prior to the launch of the Segway claimed it would 'change walking'. Similarly, in the case of coherence, for the government to fund nanotechnology research, they must 'believe' nanotechnology is a 'thing', as opposed to a name applied by some to a rather miscellaneous selection of materials science research activities. Therefore, proponents have an incentive to claim coherence where others might disagree.

These studies have been retrospective, but their data sources are contemporaneous with technology emergence so the method could be extended to contemporary analyses.  Moreover, mapping of expectations can be combined with scientometrics when suitable data become available. \cite{Gustafsson2015}, for example, used technological co-classification to identify clusters of patents, the full-text of which is subsequently analysed qualitatively to detect guiding images or \textit{leitbild}, which are generalisations shared by several actors which guide actors towards similar objectives. Guiding images are used to explain the dynamics of expectations.

Retrospective analyses can rely more extensively on scientometrics, although this has not been done very often. Scientometricians have mostly focused on the detection and analysis of growth and novelty, whereas impact seems to be taken for granted. Nonetheless, scientometrics can greatly contribute to evaluating the impact of a potentially emerging technology. A number of techniques can be used to produce intelligence on the emergence process. These include the analysis of highly-cited documents, of authorship data to generate intelligence about the actors drawn into knowledge creation processes over time (e.g. private vs. public organisations and incumbents vs. newcomers), and of changes in the collaboration structure as mapped with co-authorship data \citep[e.g.][]{Hicks1986,Melin1996}. Impact on knowledge production processes can instead be assessed by examining the dynamics of cognitive networks obtained from the study of the citations or the co-occurrence of terms across a particular set of documents.

%-------------------------------------------------------------------------------
\subsection{Uncertainty and ambiguity}
Emerging technologies are characterised by uncertainty in their possible outcomes and uses, which may be unintended and undesirable, as well as by ambiguity in the meanings different social groups associate with the given technology \citep{Stirling2007a,Mitchell2007}. For analyses of contemporary emerging technologies, news articles, editorials, review and perspective articles on professional and academic journals can be examined to qualitatively assess the degree of uncertainty and ambiguity associated with an emerging technology as well as to identify possible multiple visions of the future associated with the technology. As for the evaluation of how prominent the impact of an emerging technology will be, an STS approach to the mapping of expectations can be used for the assessment of uncertainty and ambiguity.

For retrospective analyses, the evaluation of uncertainty and ambiguity remains largely unexplored in scientometric studies, however. The few attempts made along these lines tend to overlap with those already discussed for the evaluation of the coherence attribute, since the main focus has been on the measurement of the reduction of uncertainty in scientific communication rather than on uncertainty and ambiguity associated with the potential impact or uses of emerging technologies. For example, the creation of a novel category (such as a new subject category in the classification of ISI WoS), in which subsequent journals associated with the emerging technology under examination may fall, is conceived as an indicator of increasing redundancy in the communication process --- as new journals are established and achieve a critical mass to justify the creation of a new category, the redundancy of the communication process associated with the considered emerging technology has also increased. Increasing redundancy, in turn, may indicate diminishing uncertainty.\footnote{Personal communication with Loet Leydesdorff on 2 October 2014.} In a similar vein, \cite{Lucio-Arias2009} considered words in publication titles (which are selected by authors to position knowledge claims at a given time), cited references (which enable authors to position knowledge claims in the existing socio-cognitive domain), and time as key dimensions describing the scientific discourse at the research front of a specialty. The mutual information exchanged between these dimensions (measured in terms of Shannon entropy) is suggested as an indicator of uncertainty reduction. The gap in the assessment of uncertainty and ambiguity represents, however, an important arena for future research, as we will discuss in the next section.

%-------------------------------------------------------------------------------
%	Discussion 
%-------------------------------------------------------------------------------
\section{Discussion}
We characterised emerging technologies on the basis of five attributes --- (i) radical novelty, (ii) relatively fast growth, (iii) coherence, (iv) prominent impact, and (v) uncertainty and ambiguity --- and used these to develop a framework for a coherent and systematic operationalisation of emerging technologies. A wide variety of scientometric methods are available to operationalise the various attributes of emergence. Nonetheless, these are strongly dependent on time, on the nature of the attribute, and on the data used. 

Scientometric techniques are intrinsically more effective for retrospective analyses than contemporary examinations. Time is required before documents such as publications and patents can be observed and techniques can be applied longitudinally. For example, measuring growth is particularly problematic for more contemporary analyses. Techniques using future citations are more sensitive to this issue than methods that rely on data available when documents are published (e.g.\ co-word analysis and bibliographic coupling). Lags in database indexing may also contribute to the time limitations of scientometric approaches. 

Scientometrics is also of little use in the operationalisation of uncertainty and ambiguity. The focus of scientometrics has been mainly on the detection of what is emerging, rather than on characterising the potential of what is detected to be emerging.  To our knowledge, this area is largely unexplored. Likewise, the methods reviewed in this paper show no explicit focus on how the societal aspect of prominent impact can be assessed. This is somewhat surprising when one considers the extensive scientometric work carried out for research evaluation purposes.

Furthermore, most studies have focused on publications and patents --- data that are not only sensitive to time, but also provide limited perspectives on the multifaceted phenomenon of emerging technologies. A few studies have focused on the use of news articles and big data sources (e.g.\ Google Trends). These are clearly emerging streams in scientometric and data-mining research, but so far little attention has been paid to the use of these novel data sources in the context of emerging technologies. 

The risk that detected technological emergence may be merely an artefact of the method used adds to these limitations. The reviewed methodologies rely on different models, data, thresholds, clustering algorithms and parameters, the selection of which may bias the detection of emergence towards certain patterns. For example, technological emergence is often detected with comparatively static analyses rather than with dynamic examinations. Data for a given observation period are divided into time windows and algorithms are then applied to the sample of data included in each time window. Results may vary with the number and length of time windows. Shorter time windows may not identify certain patterns of emergence because they do not capture a critical mass of documents, while longer time windows may miss cases of technologies that exhibit emerging features for a shorter period (e.g. promising technologies that eventually do not emerge). Also, the identified emerging technologies may be biased towards certain topics. \cite{Small2014}, for example, found that topics identified as emerging by the combined 'direct citation-co-citation' approach are in areas that are more likely to offer practical outcomes than non-emerging topics. This may suggest that such areas attract more resources, which, in turn, may favour the recruitment of researchers \citep{Small2014}. Yet, the identification of these emerging areas may also be the result of the model and data used. The field could move forward more confidently if instead of every study using a different data set, a standard model dataset was developed to which all techniques could be applied and the results compared \citep{Katz1996a}.

We have argued that qualitative STS approaches can be particularly powerful for overcoming the limits of scientometrics, for instance, in relation to prominent impact and to uncertainty and ambiguity. For example, mapping expectations through content analysis of news, review articles, and policy documents can provide important insights. Because STS focuses on human agency, the importance of expectations and visions in steering emergence as well as the examination of niche-regime dynamics is more apparent. Hence, this tradition attempts to address questions of how emergence happens. This may favour meaningful interpretations of scientometric data and possibly a better conceptual understanding. Scientometrics, in turn, can bring a more robust empirical approach to the STS research tradition, including the capability to address measurement error by means of statistical inference as well as to increase the generalisability of results. Few studies have followed a combined scientometrics-STS approach. \cite{Kuusi2007}, for example, applied a bibliographic coupling approach to identify clusters of patents and then to map 'guiding images' used by different actors to develop a consensus around the goals and directions during different phases of development of an emerging field.\footnote{As noted earlier in Section 4.4, a similar mixed approach has been adopted by \cite{Gustafsson2015}.} Yet, there remains great potential for substantial links and a deeper synthesis between the two traditions focusing on the examination of emergence in science and technology.

The conceptualisation and operationalisation of emerging technologies offer a number of opportunities for future research. From a conceptual point of view, more understanding of the origins of emerging technologies is required. In the early phase of emergence (high levels of radical novelty and of uncertainty and ambiguity, low levels of growth, coherence, and impact) some technologies acquire a certain momentum to become 'emerging' (when the levels of attributes are subject to more dramatic change), other technologies instead arrive at the verge of becoming emergent, but eventually not emerge at all. Funding and research programmes, the power distribution among actors, communities of practices, and regulations are likely to exert a significant impact on this process. However, more systematic research is required on the factors that enable a technology to eventually become emergent. This also extends to the empirical investigation of emergence. Studies often tend to analyse emerging technologies, without comparing them with a counter-factual sample of technologies that had the potential of becoming emergent, but eventually did not emerge. Likewise, we have limited knowledge of the end point of the emergence process, i.e.\ when emergence is over, or perhaps prematurely grinds to a halt or reverses. 

The limitations of the use of scientometrics for the operationalisation of some the attributes also represent important avenues for future research. In this regard, the use of novel data sources such as publication-full-text and funding data seems particular promising. For example, publication full-text data have been mainly used to improve the accuracy of standard scientometric approaches (e.g. co-citation and co-word clustering) \citep[e.g.][]{Boyack2013,Glenisson2005}. However, the analysis of the full-text of publications may also provide information for operationalising the uncertainty and ambiguity attribute of emergence. Instances of multiple and competing envisioned applications of an emerging technology may be identified in publication sections such as the introduction and discussion, which also have the advantage of being structured in a relatively standard manner across publications as compared to other sections. Sentiment and narrative analysis techniques may be particularly suitable for the extraction of this information.

Funding data may also provide relevant information for the operationalisation of emerging technologies. For example, uncertainty and ambiguity may be indicated by more extensive public funding than private investment. Growth in funding may indicate relatively fast growth, thus overcoming the time lag between actual emergence and emergence detected in publications and patents. The amount of funding can also cast light on the expected impact of the technology. Relatively large investments, suggest prominent impact is expected. Nonetheless, the coverage of funding data remains limited \citep{Hopkins2013}. A number of databases (e.g.\ Researchfish, FundRef, RCUK Gateway to Research, NIH RePORTER) have been recently built with aim of providing access to these data. Such databases include data on funding from major funders (e.g. government departments, research councils, large charities and foundations), but inevitably lack information on a large variety of relatively small funding organisations that may be important, especially in the early phases of development. The use of funding data as reported by authors in the acknowledgements section of publications provides better coverage of funders but no information on the amount of funding. 

The use of big data and altmetrics (e.g.\ download statistics, number of retweets, Mendeley readers, citations in blogs or news articles) add to the set of potential data sources. Given that these data are produced in a 'real-time' manner as compared to conventional scientometric data, they seem particularly promising in enabling the development of indicators for early detection. For example, publication download statistics can provide an early indication of relatively fast growth and perhaps of prominent impact in the academic domain as compared to conventional citation data. Numbers of tweets and citations in blogs or news articles may instead provide an indication of attention outside the academic domain. Nonetheless, there is first a need to improve our understanding of these data as well as how the data can be compared across different cases of emerging technologies. We hope the framework offered here can be used to structure exploration of novel data sources for the detection of emerging technology.

%-------------------------------------------------------------------------------
%	Conclusions 
%-------------------------------------------------------------------------------
\section{Conclusions}
Emerging technologies have assumed increasing relevance in the context of policy-making for their perceived ability to change the \textit{status quo} \citep[e.g.][]{Martin1995,Day2000,Alexander2012,Cozzens2010}. This has spurred \textit{ad hoc} governmental actions such as the "Future \& Emerging Technologies" (FET) initiative funded by the European Commission in 2013 and the "Foresight and Understanding from Scientific Exposition" (FUSE) research program funded by the US Intelligence Advanced Research Projects Activities (IARPA) in 2011. The FUSE program, for example, in pursuit of potential uses of big data, has aimed to develop methods for the reliable early detection of emergence in science and technology by mining the full-text of publications and patents. Policy interest has been matched by the academic community who have developed a variety of methods for the detection and analysis of technological emergence in recent years especially in the scientometric domain \citep[e.g.][]{Small2014,Glanzel2012}.

Despite this broad interest, a widely accepted definition of emerging technologies and an agreed conceptually grounded framework for their operationalisation are both still missing. We showed that emerging technologies are either loosely defined in the empirical literature or often no definition at all is provided. As a consequence, operationalisations of emergence tend to differ greatly even between approaches using the same techniques. In addition, the understanding of what is an emerging technology differs across actors: some individuals may conceive a technology to be emergent because they expect impact on the socio-economic system, while others may see the same technology as old and no longer emergent. This, in turn, has significant implications for policy making and the governance of emerging technologies.

The present paper has attempted to move the field forward by systematically delineating the concept of technological emergence linked to measurement options. To do so, we first developed a definition of emerging technologies that is able to capture the multifaceted nature of emerging technologies, and then proposed a framework for their operationalisation drawing on, but not limited to, scientometric analysis. We identified five attributes of emerging technologies: (i) radical novelty, (ii) relatively fast growth, (iii) coherence, (iv) prominent impact, and (v) uncertainty and ambiguity, and defined emerging technologies as: "\textit{a relatively fast growing and radically novel technology characterised by a certain degree of coherence persisting over time and with the potential to exert a considerable impact on the socio-economic domain(s) which is observed in terms of the composition of actors, institutions and the patterns of interactions among those, along with the associated knowledge production processes. Its most prominent impact, however, lies in the future and so in the emergence phase is still somewhat uncertain and ambiguous}". 

We then developed a coherent and systematic framework for operationalising these attributes of emergence. Scientometric literature was the main source of potential measures. Relevant studies were reviewed and linked to the attributes of emergence.  Our analysis showed that scientometric analysis is particularly appropriate for the operationalisation of growth, novelty and coherence.  Relatively fast growth is operationalised in many studies and often evaluated by counting documents over time (such as news articles, publications, and patents) \citep[e.g.][]{Porter1995}. Radical novelty is identified with the appearance of new clusters of documents or words in citation or co-word analyses \citep[e.g.][]{ Kajikawa2008}, while other studies point to the importance of also considering the extent to which the new cluster is connected to clusters in the same year of observation or to clusters identified in previous years \citep[e.g.][]{Small2014}. Indicators based on entropy measures or on the appearance of new categories (e.g. journals, technological classes, terms in institutionalised vocabularies) were identified as more suitable for assessing coherence \citep[e.g.][]{Cozzens2010}.

Nonetheless, important limitations exist on the scientometric contribution to the operationalisation of emerging technologies. The evaluation of uncertainty and ambiguity as well as prominent impact is, for example, largely unexplored in scientometrics. Also, methods often rely on few data sources, mostly publication and patent data, which tend to be not suitable for the analysis of contemporary cases of emerging technologies --- these data require time to be generated. The risk that detecting apparent emergence may be merely an artefact of selected models adds to these limitations. 

We have argued that the qualitative investigation of emerging technologies conducted by STS researchers seems particularly promising in complementing scientometrics for the purpose of operationalising the attributes of emergence. The mapping of expectations of emerging technologies by mean of qualitative analysis of documents such as news, review articles, and policy documents can, for example, provide important insights on the uncertainty and ambiguity and the prominent impact attributes of emergence, especially in the case of contemporary analyses. STS approaches can also provide meaningful interpretation of the results of scientometrics, thus potentially reducing the likelihood of detecting false positives or missing patterns.

We envisage a number of opportunities for future research.  First, future research should pay more attention to the origins of emerging technologies. We have limited knowledge of factors that enable certain technologies to become emergent while other do not emerge at all. This also extends to the research design used for the investigation of emerging technologies. Studies often examine emerging technologies without delineating a counter-factual sample of technologies that did not emerge but which nevertheless had the potential to emerge. Similarly, we have limited knowledge on when a technology ceases to be emergent and what factors shape this process. Second, the increasing access to publication full-text, funding data, altmetrics, and, more generally big data, may provide significant opportunities for future research in scientometrics to develop indicators and methods for the evaluation of attributes of emergence for which the current 'state of the art' in scientometrics can provide only a limited contribution. 

In summary, we have showed that considerable disagreement exists on what is technological emergence and how it should be operationalised. This has important implications for policy-making in the context of emerging technologies (e.g.\ resource allocation, creation of research programmes, drawing up of regulations), which, in turn, exerts a direct effect on the emergence process itself. The present paper has attempted to contribute to this ongoing and urgent debate in science policy research through conceptual clarification of the phenomenon of emergence. This is a necessary precondition for a coherent and systematic operationalisation of emerging technologies, for future research developments, for a better understanding of the phenomenon, and, therefore for more informed policy-making and governance of emerging technologies.

%-------------------------------------------------------------------------------
%	Acknowledgements
%-------------------------------------------------------------------------------
\section*{Acknowledgements}
We acknowledge the support of the People Programme (Marie Curie Actions) of the European Union's Seventh Framework Programme (FP7/2007-2013) (award PIOF-GA-2012-331107 - \href{http://www.danielerotolo.com/netgenesis}{"{\color{blue}NET-GENESIS: Network Micro-Dynamics in Emerging Technologies}}"). We are grateful to Loet Leydesdorff, Nils Newman, Alan Porter, Andrew Stirling, Jan Youtie, the two anonymous referees of the SPRU Working Paper Series (SWPS), and the two anonymous referees of Research Policy for their comments, criticisms and suggestions. A previous version of this paper was presented at the SPRU Wednesday Seminar Series at the University of Sussex (27 May 2015), the Technology Policy Assessment Centre (TPAC) Seminar Series at the School of Public Policy of the Georgia Institute of Technology (4 December 2014), and the 2015 Eu-SPRI Conference (9-12 June 2015, Helsinki, Finland).

%-------------------------------------------------------------------------------
%	References
%-------------------------------------------------------------------------------
\newpage
\singlespace
\bibliographystyle{apalike}
\bibliography{/Users/daniele/Dropbox/References/bibtex_references/library.bib}

%-------------------------------------------------------------------------------
%	Appendix
%-------------------------------------------------------------------------------
\newpage
\section*{Appendix}
\setcounter{table}{0}
\renewcommand{\thetable}{A\arabic{table}}

%=====
\setlength{\tabcolsep}{10pt}
\renewcommand{\arraystretch}{1.5}
\begin{table}[h]\footnotesize
	\caption{\label{tab:emergence}The concept of 'emergence' in complex systems theory (studies are ordered chronologically).}
	\centering
{\begin{tabular}{C{2cm}p{13.5cm}}
\hline\hline
\textbf{Study} & \textbf{Definition}\\
\hline
\cite{Bedau1997} & "[...] Emergent phenomena are somehow constituted by, and generated from, underlying processes [...] are somehow autonomous from underlying processes" (p. 375) 
"[...] there is a system, call it S, composed out of micro level parts [...] S has various macro level states (macrostates) and various micro level states (microstates) [...] there is a microdynamic, call it D, which governs the time evolution of S's microstates [...] I define weak emergence as follows: Macrostate P of S with microdynamic D is weakly emergent iff P can be derived from D and S's external conditions but only by simulation" (p. 377-378)\\
\cite{Goldstein1999} & ''Emergence [...] as the arising of novel and coherent structures, patterns, and properties during the process of self-organization in complex systems [...] common properties that identify them as emergent: 
\begin{itemize}
\item Radical novelty: emergents have features that are not previously observed in the complex system under observation [...]
\item Coherence or correlation: emergents appear as integrated wholes that tend to maintain some sense of identity over time. This coherence spans and correlates the separate lower-level components into a higher-level unity.
\item Global or macro level: [...] the locus of emergent phenomena occurs at a global or macro level [...]
\item Dynamical: emergent phenomena are not pre-given wholes but arise as a complex system evolves over time [...]
\item Ostensive: emergents are recognized by showing themselves, i.e.\, they are ostensively recognized [...]'' (p. 49-50)
\end{itemize}\\
\cite{Corning2002} & ''Emergent phenomena be defined as a subset of the vast (and still expanding) universe of cooperative interactions that produce synergistic effects of various kinds, both in nature and in human societies [...] all emergent phenomena produce synergistic effects, but many synergies do not entail emergence. In other words, emergent effects would be associated specifically with contexts in which constituent parts with different properties are modified, reshaped, or transformed by their participation in the whole.'' (p. 23-24)\\
\cite{Chalmers2006}	& "a high-level phenomenon is strongly emergent with respect to a low-level domain when the high-level phenomenon arises from the low-level domain, but truths concerning that phenomenon are not deducible even in principle from truths in the low-level domain [...] a high-level phenomenon is weakly emergent with respect to a low-level domain when the high-level phenomenon arises from the low-level domain, but truths concerning that phenomenon are unexpected given the principles governing the low-level domain." (p. 244)\\
\cite{DeHaan2006} & "Emergence is about the properties of wholes compared to those of their parts, about systems having properties that their objects in isolation do not have. Emergence is also about the interactions between the objects that cause the coming into being of those properties, in short the mechanisms producing novelty." (p. 294)\\
\hline\hline
\multicolumn{2}{l}{\footnotesize \textit{Source: search performed by authors on Google Scholar and SCOPUS and extended to cited references.}}
\end{tabular}
}
\end{table}
%=====

%-------------------------------------------------------------------------------
\end{document}